\documentclass[prd,nofootinbib,onecolumn,superscriptaddress]{revtex4}

\usepackage{amsmath, amssymb, amsthm, graphicx, epsfig, fancyhdr,epsfig, slashed, mathrsfs,bm}

\usepackage{natbib}
\usepackage{float}
\usepackage{layout}

\usepackage{mathtools} 

\usepackage{tikz,xcolor,hyperref}

\definecolor{lime}{HTML}{A6CE39}
\DeclareRobustCommand{\orcidicon}{
	\begin{tikzpicture}
	\draw[lime, fill=lime] (0,0) 
	circle [radius=0.2] 
	node[white] {{\fontfamily{qag}\selectfont \tiny ID}};
	\draw[white, fill=white] (-0.0625,0.095) 
	circle [radius=0.007];
	\end{tikzpicture}
	\hspace{-2mm}
}

\foreach \x in {A, ..., Z}{\expandafter\xdef\csname orcid\x\endcsname{\noexpand\href{https://orcid.org/\csname orcidauthor\x\endcsname}
			{\noexpand\orcidicon}}
}

\newcommand{\be}{\begin{equation}}
\newcommand{\ee}{\end{equation}}
\newcommand{\bea}{\begin{eqnarray}}
\newcommand{\eea}{\end{eqnarray}}

\def\hhref#1{\href{http://arxiv.org/abs/#1}{arXiv:#1}} 

\begin{document}


\title{Mass gap in non-perturbative quadratic $\mathcal{R}^2$ gravity via Dyson-Schwinger}

\author{Sayantan Choudhury\orcidC{}}
\email{sayanphysicsisi@gmail.com,schoudhury@fuw.edu.pl}
\affiliation{Institute of Theoretical Physics, Faculty of Physics, University of Warsaw, ul. Pasteura 5, 02-093 Warsaw, Poland}

\author{Marco Frasca\orcidA{}}
\email{marcofrasca@mclink.it }
\affiliation{Rome, Italy}

\author{Anish Ghoshal\orcidB{}}
\email{anish.ghoshal@fuw.edu.pl}
\affiliation{Institute of Theoretical Physics, Faculty of Physics, University of Warsaw, ul. Pasteura 5, 02-093 Warsaw, Poland}

\begin{abstract}
\textit{We apply in a simple model derived from quadratic $\mathcal{R}^2$ gravity the technique of Dyson-Schwinger equations to solve for its corresponding quantum theory. Particularly, we solve the classical equations of motion to get a solution to the hierarchy of Dyson-Schwinger equations in the limit of large 
Ricci scalar, assumed to be constant and larger than the square of the Starobinsky mass. 
%
Moving to the Einstein frame,
the model admits Higgs-like solutions with a single particle having a finite mass. 
We quantize the scalar field showing the appearing of a mass gap through a Higgs-like solution. The presence of the mass gap, that increases with the square root of the Ricci scalar, shows how the effect of the scalar sector at low-energy becomes ineffective, making it relevant only at short distances.
}
\end{abstract}

\maketitle


\section{Introduction}

One of the most successful theories of the past century and certainly the most successful theory in the context of gravity is Albert Einstein's general relativity (GR). This theory has stood the test of time again and again which has been verified by several observations. These experimental observations range from those as recent as the discovery of gravitational waves, with coalescing binary black hole astrophysical systems~\cite{LIGOScientific:2016aoc}, to the extraordinary optical images of the black holes in the center of our galaxy and the M87 one as captured by the Event Horizon Telescope~\cite{EventHorizonTelescope:2019dse,EventHorizonTelescope:2019ggy,EventHorizonTelescope:2022xnr,EventHorizonTelescope:2022wok,EventHorizonTelescope:2022xqj}. Extending GR, the theory suitably can account for matter fields, and forms the edifice of our present understanding of modern day cosmology, including the early universe right from the origin of hot big bang.

Even with so much success at large scales, nonetheless, at the quantum frontier GR is known to be non-renormalizable by perturbative methods~\cite{Goroff:1985sz,Goroff:1985th} and may not remain perturbative at length scales much below the Planck one. Also it encounters Big Bang and Black Hole singularity problems and known to pathological behaviour in the ultra-violate (UV) \cite{Salvio:2018crh}.
GR also suffers from the what is known as the ``conformal-factor problem": in simple terms this is nothing but absence of a consistent Euclidean path integral. Moreover this becomes quite serious since all known quantum field theories (QFT) can be consistently defined only as analytic continuations of the corresponding Euclidean version, see Ref. \cite{Salvio:2024joi} for details.

The issue of non-renormalizability and the issue of conformal-factor is solved when GR is augmented with quadratic-in-curvature terms which is often called quadratic gravity (See~\cite{Salvio:2018crh,Donoghue:2021cza} for reviews.). It has been shown to be renormalizable~\cite{Weinberg:1974tw,Deser:1975nv,Stelle:1976gc,Barvinsky:2017zlx} and following the prescription of Ref.~\cite{Gibbons:1994cg} one may determine the Euclidean action which implies that it is also free from the conformal-factor problem in most of the regions of its parameter space~\cite{Menotti:1989ms}.
Such quadratic gravity theories also address the singularity issues. 
The classical Hamiltonian of quadratic gravity is unbounded from below, something known in the literature as the Ostrogradsky's instability issue as shown in Ref.~\cite{ostro}. Frankly, this is common in all kinds of higher-derivative gravity extensions. Nonetheless estimations of runaway rates show that such a decay can be slow enough to be compatible with observations as studied in Refs.~\cite{Salvio:2019ewf,Gross:2020tph,Held:2021pht,Held:2023aap}. 


In this paper we offer a non-perturbative formulation of quadratic $\mathcal{R}^2$ gravity. This is based on the Dyson-Schwinger approach and may certainly provide us with a fully quantum and non-perturbative justification for these classical cosmological findings in the long run. Furthermore, such a formulation that we show here can also be utilised to validate and check consistency with other independent perturbative methods and sometimes even background-dependent approaches to formulate quadratic gravity which been shown to preserve  unitarity and (meta)stablity. In the context of renormalizable theories, these approaches include the Lee-Wick~\cite{Lee:1969fy,Salvio:2018kwh,Donoghue:2019fcb,Holdom:2021hlo} the fakeon prescription~\cite{Anselmi:2017ygm} and  string-inspired non-local ghost-free approaches to higher-derivative gravity \cite{Frasca:2020jbe,Frasca:2020ojd,Frasca:2021iip, Frasca:2022duz,Frasca:2022gdz}. Particularly for quadratic gravity this has been studied rigorously in Ref. \cite{Salvio:2024joi}.

The novel tool to treat the non-perturbative systems in the framework of Dyson-Schwinger Equations that we propose in this paper are based on the exact solution of the background equations of motion in terms of Jacobi elliptic functions. This is inspired by the analytic approach of Dyson-Schwinger equations (DSE), originally devised by Bender, Milton and Savage~\cite{Bender:1999ek}. Using this we are able to represent the Green's functions of the theory purely analytically, therefore we understand the effect of the background on the interactions that remains to be valid even in the strong coupling regimes~\cite{Frasca:2015yva}. This approach has been widely used to investigate QCD with great success with respect to results obtain closely matching with lattice QCD results~\cite{Frasca:2021yuu,Frasca:2021mhi,Frasca:2022lwp,%
Frasca:2022pjf,Chaichian:2018cyv} and to the scalar
sector~\cite{Frasca:2015wva}\footnote{Moreover extensions beyond the traditional QFT and SM of particle physics to other types of
models involving new gauge sectors and string-inspired non-local
theories have been widely discussed using this methodology~\cite{Frasca:2019ysi,Chaichian:2018cyv,Frasca:2017slg,Frasca:2016sky,Frasca:2015yva,Frasca:2015wva,Frasca:2013tma,Frasca:2012ne,Frasca:2009bc,Frasca:2010ce,Frasca:2008tg,Frasca:2009yp,Frasca:2008zp,Frasca:2007uz,Frasca:2006yx,Frasca:2005sx,Frasca:2005mv,Frasca:2005fs}}. In context to experiments concerning particle physics phenomenology like non-perturbative hadronic vacuum polarization contributions to the muon anomalous magnetic moment $(g-2)_\mu$ was predicted in Ref.\cite{Frasca:2021yuu}, QCD in the non-perturbative regimes~\cite{Frasca:2021mhi,Frasca:2022lwp,Frasca:2022pjf}, Higgs-Yukawa theory \cite{Frasca:2023qii}, finite temperature field theory \cite{Frasca:2023eoj},
non-perturbative false vacuum transitions leading to phase transitions~\cite{Frasca:2022kfy,Calcagni:2022tls,Calcagni:2022gac}, dark energy \cite{Frasca:2022vvp}, and investigations
of the mass gap and confinement in string-inspired infinite-derivative and Lee-Wick theories~\cite{Frasca:2020jbe,Frasca:2020ojd,Frasca:2021iip, Frasca:2022duz,Frasca:2022gdz}. These tools were also employed to implement dynamical generation electroweak (EW) scale via scalar, vector boson and fermionic condensates too very recently \cite{Frasca:2024fuq,Chatterjee:2024dgw,Frasca:2024pmv,Frasca:2023eoj,Frasca:2024ame} as an approach to address the gauge hierarchy problem in the Standard Model (SM) of particle physics.

 A final formulation of the theory should involve quantum cosmology with a quantum mechanical ``wave function of the universe", as investigated by Hawking and Hartle~\cite{Hartle:1983ai}. This actually can be used to solve the Schrodinger equation, also known as Wheeler-DeWitt equation in this context~\cite{DeWitt:1967yk,Wheeler}. Therefore a fully consistent quantum cosmology with GR extended to quadratic gravity is what we allude to, since this theory contains Starobinsky's model of inflation (also known as $\mathcal{R}^2$ inflation)~\cite{Starobinsky:1980te}, which is currently in perfect agreement with the cosmic microwave background radiation (CMBR) observations assuming the cosmic inflationary paradigm~\cite{Ade:2015lrj,Planck2018:inflation,BICEP:2021xfz}.
 
 Yet another application of a non-perturbative quadratic gravity via Dyson-Schwinger approach would be in the paradigm of asymptotic safety originally proposed by Steve Weinberg~\cite{WeinbergAS}, where gravity is made ultra-violate (UV) complete and shown to be renormalizable due to presence of a non-perturbative UV fixed point. Investigations of such fixed points have been extensively carried out for quadratic $\mathcal{R}^2$ gravity in Refs.~\cite{Benedetti:2009rx,Falls:2020qhj} which lead to consistent realization of Asymptotic Safe Gravity.

 As we will show that the Einstein-Hilbert term plus a quadratic term of the Ricci scalar, for strong fields. Conformal symmetry is broken by the appearance of a mass gap in the theory. The mass spectrum of theory is a discrete spectrum with an infinite tower of harmonic-oscillator like excitations. The mass gap scale is proportional to the Starobinsky mass which is the coefficient of the $\mathcal{R} ^2$ term which represents the self-interactions among the gravitons and the scalar degrees of freedom.

In this work, we start considering generally non-renormalizable theories and obtain some meaningful renormalizable scalar theories to work with, only under some conditions. There exists several such examples, a well-known example being given by the Nambu-Jona-Lasinio model used to describe strong interactions \cite{Klevansky:1992qe}. This model is characterized by fermion fields with a quartic interaction between them. As such, this model is not renormalizable and the computations depend on a cut-off. When a bosonization procedure is applied \cite{Klevansky:1992qe}, a $\sigma$-model emerges that is renormalizable, and could represent a description of the breaking of the chiral symmetry in strong interactions. With this principle in mind, we try to have similar approach here showing that, under proper conditions, some effective scalar models could be extracted from quadratic gravity and can be suitably properly studied using the well-known techniques of quantum field theory as the Dyson-Schwinger equations are. However a concrete re-normalizable model of gravity like Stelle gravity or asymptotic safe gravity or non-local gravity (see Ref. \cite{Salvio:2018crh} for such a discussion) is beyond the scope of the current paper and will be studied in future.

%



The paper is organized as follows: In Sec.~\ref{sec2}, we introduce the model and derive the scalar field theory arising from it. In Sec.~\ref{sec3}, we discuss the classical solutions for the scalar field theory. In Sec.~\ref{sec4}, we quantize the model ans solve the set of Dyson-Schwinger equations both with a null or non-null the cosmological constant. In Sec.\ref{sec5}, we provide the conclusions.

\medskip

\section{Scalar Model for Quadratic Starobinsky gravity: A small review\label{sec2}}

We present a scalar field model from the full action of the quadratic gravity \cite{Starobinsky:1980te,Mukhanov:1981xt,Kaneda:2010ut} and then we will quantize it with the technique of the Dyson-Schwinger equations\footnote{We do not consider here a scenario like Higgs inflation, involving a term like $\xi \phi^2 \mathcal{R}$. We only restrict ourselves to a minimally coupled scalar field to gravity as comes out from quadratic gravity.}.
\be
\label{eq:star}
S=-\frac{M_p^2}{2}\int d^4x\sqrt{-g}\left(\mathcal{R}-2\Lambda+\frac{\mathcal{R}^2}{6 M^2}\right),
\ee
where $M_p$ is the Planck mass and $M$ is a new mass scale in this theory which is called the Starobinsky mass or also known as the conformal coupling. This action can be re-written in terms of the interaction between gravity and a scalar field $\chi$ after performing the transformation \cite{Whitt:1984pd,Kehagias:2013mya}
\be
g_{\mu\nu}\rightarrow e^{-\sqrt{\frac{2}{3}}\frac{\chi}{M_{\rm p}}}g_{\mu\nu},
\ee
where $g_{\mu\nu}$ is a generic metric.
%
This gives \cite{Salvio:2018crh}:
\be
  S =\int {\rm d}^4 x\sqrt{-g}\left[-\frac{M_{\rm p}^2}{2}(\mathcal{R}-2\Lambda e^{-2\sqrt{\frac{2}{3}} \frac{\chi}{M_p}})+\frac{1}{2}g^{\mu\nu}\partial_\mu\chi\partial_\nu \chi+\frac{3}{4} M_{\rm p}^2 M^2\left(1-e^{-\sqrt{\frac{2}{3}}\frac{\chi}{M_{\rm p}}}\right)^2\right]. \label{R3}
\ee
%
This action yields the following sets of equations of motion, which are given by $\frac{\delta S}{\delta g_{\mu \nu}}$ and $\frac{\delta S}{\delta \chi}$:


\bea
&&\mathcal{R}_{\mu\nu} = \frac{1}{M_p^2} \left\{ 
 \partial_\mu \chi \, \partial_\nu \chi
- \tfrac{1}{2} g_{\mu\nu} \, g^{\alpha\beta} \partial_\alpha \chi \, \partial_\beta \chi
- g_{\mu\nu} \left[
M_{\rm p}^2 \Lambda \, e^{-2\sqrt{\tfrac{2}{3}} \tfrac{\chi}{M_{\rm p}}}
+ \tfrac{3}{4} M_{\rm p}^2 M^2 \left(1 - e^{-\sqrt{\tfrac{2}{3}} \tfrac{\chi}{M_{\rm p}}}\right)^2
\right]
\right\}
+\frac{1}{2}g_{\mu\nu}{\cal R},
\nonumber \\
&&\Box_g\chi= \sqrt{\frac{3}{2}} M^2 M_p e^{-\sqrt{\frac{2}{3}} \frac{\chi}{M_p}} \left(1 - e^{-\sqrt{\frac{2}{3}} \frac{\chi}{M_p}}\right) - \sqrt{\frac{8}{3}} \Lambda M_p e^{-2\sqrt{\frac{2}{3}} \frac{\chi}{M_p}}
\eea
where $\Box_g=-\frac{1}{\sqrt{-g}}\partial_\mu g^{\mu\nu}\sqrt{-g}\partial_\nu$ is the Laplace-Beltrami operator.

We look for solution with $R$ being a constant. This yield the set of equations for the scalar field
\bea
&&\partial_\sigma\chi\partial^\sigma\chi+3M_{\rm p}^2 M^2\left(1-e^{-\sqrt{\frac{2}{3}}\frac{\chi}{M_{\rm p}}}\right)^2+4M_p^2\Lambda e^{-2\sqrt{\frac{2}{3}} \frac{\chi}{M_p}}-M_p^2\mathcal{R}=0, \nonumber \\
&&-\frac{1}{\sqrt{-g}}\partial_\mu g^{\mu\nu}\sqrt{-g}\partial_\nu\chi=\sqrt{\frac{3}{2}} M_{\rm p} M^2\left(1-e^{-\sqrt{\frac{2}{3}}\frac{\chi}{M_{\rm p}}}\right)
e^{-\sqrt{\frac{2}{3}}\frac{\chi}{M_{\rm p}}}-\sqrt{\frac{8}{3}} \Lambda M_p e^{-2\sqrt{\frac{2}{3}} \frac{\chi}{M_p}}.
\eea
Taking a dimensionless rescaled field variable $\varphi=\sqrt{\frac{2}{3}}\frac{\chi}{M_{\rm p}}$, 
we further get:
\bea
\label{eq:fullset1}
&&\partial_\sigma\varphi\partial^\sigma\varphi+ 2M^2\left(1-e^{-\varphi}\right)^2+\frac{8}{3}\Lambda e^{-2\varphi}-\frac{2}{3}\mathcal{R}=0 \nonumber \\
&&\Box_g\varphi=M^2\left(1-e^{-\varphi}\right)e^{-\varphi}-\frac{4}{3}\Lambda e^{-2\varphi}.
\eea
Let us take $\varphi=-\log Z$, we get 
\bea
\label{eq:fullset2}
&&Z^{-2}\partial_\sigma Z\partial^\sigma Z+ 2M^2\left(1-Z\right)^2+\frac{8}{3}\Lambda Z^2-\frac{2}{3}\mathcal{R}=0 \nonumber \\
&&-Z^{-1}\Box_g Z+Z^{-2}\partial_\sigma Z\partial^\sigma Z=M^2\left(1-Z\right)Z-\frac{4}{3}\Lambda Z^2.
\eea
We insert the first equation into the second obtaining
\be
\Box_g Z=-\left(2M^2-\frac{2}{3}\mathcal{R}\right)Z+3M^2Z^2-\left(M^2+\frac{4}{3}\Lambda\right) Z^3.
\ee
Thus, we are left with a quartic scalar $\phi^4$-like theory which can be easily seen upon the introduction of the following mapping of variables: 
\be
\label{eq:resc}
Z=\frac{M^2}{M^2+\frac{4}{3}\Lambda}+\frac{\phi}{\sqrt{M^2+\frac{4}{3}\Lambda}},
\ee
that gives us
\bea
\label{eq:eom}
&&\Box_g\phi=-\left(2M^2-\frac{2}{3}\mathcal{R}\right)\frac{M^2}{\sqrt{M^2+\frac{4}{3}\Lambda}}+\frac{2M^6}{\left(M^2+\frac{4}{3}\Lambda\right)^\frac{3}{2}} \nonumber \\
&&+\left[-\left(2M^2-\frac{2}{3}\mathcal{R}\right)+\frac{3M^4}{M^2+\frac{4}{3}\Lambda}\right]\phi-\phi^3
\eea
Such a potential could signal a false vacuum decay yielding, after integrating on $\phi$ and changed sign to the rhs of eq.(\ref{eq:eom}),
\be
\label{eq:Vfvd}
V(\phi)=\left[\left(2M^2-\frac{2}{3}\mathcal{R}\right)\frac{M^2}{\sqrt{M^2+\frac{4}{3}\Lambda}}-\frac{2M^6}{\left(M^2+\frac{4}{3}\Lambda\right)^\frac{3}{2}}\right]\phi -\frac{1}{2}\left[-\left(2M^2-\frac{2}{3}\mathcal{R}\right)+\frac{3M^4}{M^2+\frac{4}{3}\Lambda}\right]\phi^2+\frac{1}{4}\phi^4.
\ee
We have set the integration constant to 0.
\begin{figure}[H]\begin{center}
\includegraphics[scale=0.45]{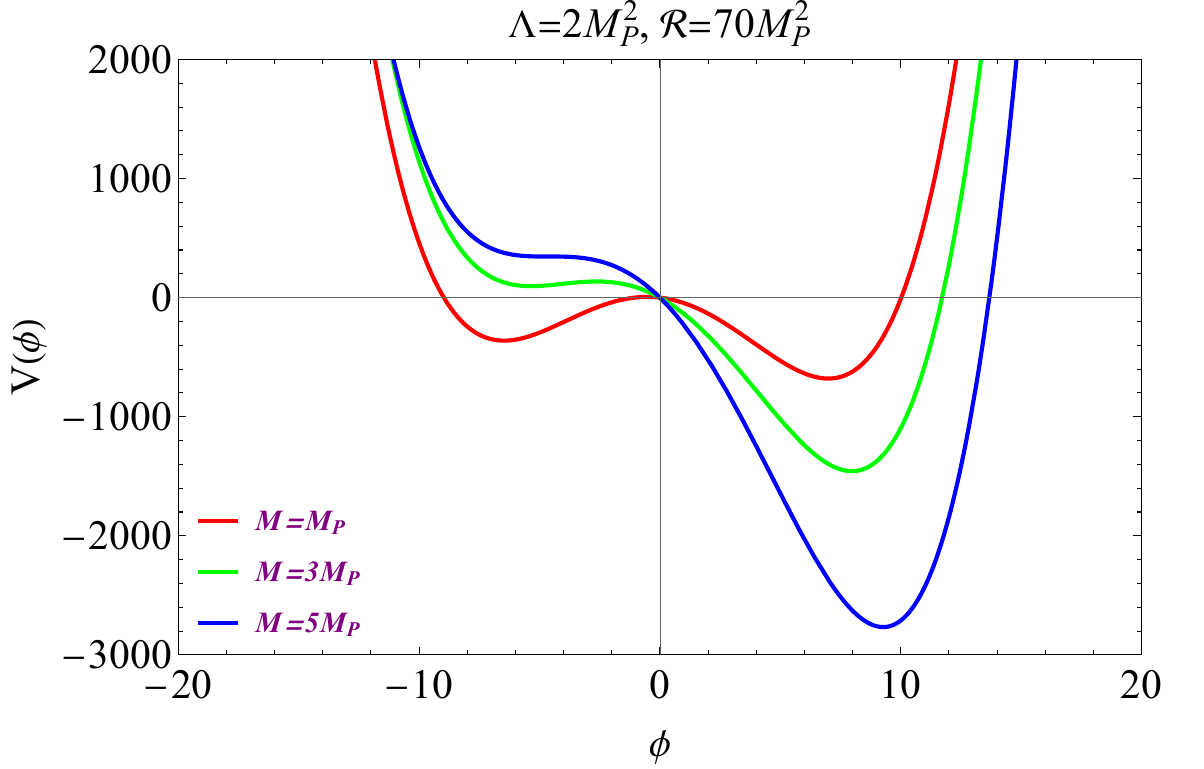}
\caption{\label{fig1} \it Plot of the potential of the conformal factor. It is seen the possibility of false vacuum decay for 
$3M^2<\mathcal{R}$.
All the physical quantity are dimensionless being scaled with the Planck mass $M_P$.}
\end{center}\end{figure}
Fig.~\ref{fig1} shows different potentials varying the Starobinsky mass $M$ with respect to the Planck mass for a given value of the cosmological constant. The characteristic shape of a false vacuum decay potential can be recognized
provided the condition $3M^2<\mathcal{R}$ holds.
%

The condition $3M^2<\mathcal{R}$ (strong coupling regime) is essential for our analysis. It grants the reduction of the Laplace-Beltrami operator to that of the flat space, aside from a proper scale changing. Physically, this means that the quadratic term $\mathcal{R}^2$ is prevailing with respect to the linear one of the standard Einstein-Hilbert action. For this reason, the quantum behavior in this regime is the main concern of the present study.

We see that the presence of the cosmological constant is not so relevant and so, in the following we set it to 0. This will yield the equation of motion
\be
\label{eq:eom_1}
\Box_g\phi=\frac{2}{3}\mathcal{R}M+\left[\frac{2}{3}\mathcal{R}+M^2\right]\phi-\phi^3,
\ee
and the potential
\be
V(\phi)=-\frac{2}{3}\mathcal{R}M\phi-\frac{1}{2}\left[\frac{2}{3}\mathcal{R}+M^2\right]\phi^2+\frac{\phi^4}{4}.
\ee

\medskip

\section{Classical solutions\label{sec3}}

For our study, 
we need to introduce an arbitrary source $j$ and compute the functional series
\be
\phi(x)=\phi_0(x)+\sum_{n=1}^\infty\int C_n(x,x_1,\ldots,x_{n-1})\prod_{k=1}^nj(x_k)d^4x_k,
\ee
where we are assuming $\phi$ a functional of an arbitrary source $j$ and
\be
C_n(x,x_1,\ldots,x_{n-1})=\left.\frac{\delta^n\phi(x)}{\delta j(x_1)\ldots\delta j(x_{n-1})}\right|_{j=0}.
\ee
By iterating the derivation with respect to $j$ and starting from the equation of motion (that holds for $3M^2<\mathcal{R}$),
\be
\Box\phi(x)=\Omega+\mu_0^2\phi(x)-\phi^3(x)+j(x),
\ee
where 
$\Omega=\frac{2}{3}\mathcal{R}M$ and $\mu_0^2=\frac{2}{3}\mathcal{R}+M^2=\frac{\Omega}{M}+M^2$ having set $\Lambda=0$ to make handling the equations simpler,
we get
\bea
\Box\phi_0(x)&=&\Omega+\mu_0^2\phi_0(x)-\phi_0^3(x), \nonumber \\
\Box C_1(x,x_1)&=&\left[\mu_0^2-3\phi_0^2(x)\right]C_1(x,x_1)+\delta^4(x-x_1), \nonumber \\
\Box C_2(x,x_1,x_2)&=&\left[\mu_0^2-3\phi_0^2(x)\right]C_2(x,x_1,x_2)-6\phi_0(x)C_1(x,x_2)C_1(x,x_1),
\nonumber \\
\Box C_3(x,x_1,x_2,x_3)&=&\left[\mu_0^2-3\phi_0^2(x)\right]C_3(x,x_1,x_2,x_3)
-6C_1(x,x_1)C_1(x,x_2)C_1(x,x_3)-6\phi_0(x)C_2(x,x_1,x_2)C_1(x,x_3) \nonumber \\
&&-6\phi_0(x)C_1(x,x_1)C_2(x,x_2,x_3)-6\phi_0(x)C_2(x,x_1,x_3)C_1(x,x_2), \nonumber \\
&\vdots&.
\eea
The cubic polynomial on the lhs of the leading order equation has three solution as expected $(-M,(3M-\sqrt{3}\sqrt{3M^2+8R})/6,(3M+\sqrt{3}\sqrt{3M^2+8R})/6)$. The last one is the 
true
vacuum. 
Thus, if we select it, we can work out a Higgs-like solution and a massive solution already at this stage. We will discuss this point below. On the other hand, we can exploit the strong coupling condition $3M^2<\mathcal{R}$ and neglect the constant $\Omega$ with respect to the other terms in the leading order equation. We can also approximate $\mu_0^2\approx 2R/3$ that can be seen as a leading order approximation\footnote{We observe that we are claiming that $RM\ll R\phi_0$ that is $\phi_0\gg M$ that is consistent with our approximation discussed in the previous section.}. This implies that the leading order equation takes the form
\be
\Box\phi_0(x)\approx\mu_0^2\phi_0(x)-\phi_0^3(x),
\ee
being $\mu_0^2$ the approximate value given above.
In this way, we are able to get an alternative approximate solution to
this set of equations yielding
\be
\phi_0(x)=\sqrt{\frac{2}{3}}\mu_0\text{dn}\left(p\cdot x+\theta,-1\right),
\ee
where one can see M as the Starobinsky mass in the original Lagrangian (\ref{eq:star}), dn is a Jacobian elliptic function and
\be
p^2=\frac{\mu_0^2}{3}.
\ee
Similarly, also in this case 
\be
\label{eq:Gom1}
C_1(p)=\frac{\sqrt{2}\pi^3}{K^3(-1)}\sum_{n=1}^\infty n^2\frac{e^{-n\pi}}{1+e^{-2n\pi}}\frac{1}{p^2+m_n^2},
\ee
where $K(-1)$ is the complete elliptic integral of the first kind and
\be
\label{eq:spe1}
    m_n=n\frac{\pi}{K(-1)}\frac{\mu_0}{\sqrt{3}}.
\ee
We can iterate obtaining
\bea
C_2(x,x_1,x_2)&=&-6\int d^4x'C_1(x,x')\phi_0(x')C_1(x',x_2)C_1(x',x_1), \nonumber \\
C_3(x,x_1,x_2,x_3)&=&-6\int d^4x'C_1(x,x')C_1(x',x_1)C_1(x',x_2)C_1(x',x_3)-6\int d^4x'C_1(x,x')\phi_0(x')C_2(x',x_1,x_2)C_1(x',x_3) \nonumber \\
&&-6\int d^4x'C_1(x,x')\phi_0(x')C_1(x',x_1)C_2(x',x_2,x_3)-6\int d^4x' C_1(x,x')\phi_0(x')C_2(x',x_1,x_3)C_1(x',x_2),  \nonumber \\
&\vdots&.
\eea
We see that our solution is completely obtained by knowing $\phi_0$ and the Green function $C_1(x,x')$ that is translation invariant and can be written as $C_1(x-x')$. We will see that the same result can be obtained in quantum field theory providing a Gaussian solution to the set of Dyson-Schwinger equations of the theory.

The Higgs-like case can analyzed by taking
\be
\phi_0\approx \mu_0.
\ee
This reduces the set of equations to
\bea
\Box C_1(x,x_1)&=&-2\mu_0^2C_1(x,x_1)+\delta^4(x-x_1), \nonumber \\
\Box C_2(x,x_1,x_2)&=&-2\mu_0^2C_2(x,x_1,x_2)-\frac{6}{\sqrt{2}}MC_1(x,x_2)C_1(x,x_1),\nonumber \\
\Box C_3(x,x_1,x_2,x_3)&=&-2\mu_0^2C_3(x,x_1,x_2,x_3)
-6C_1(x,x_1)C_1(x,x_2)C_1(x,x_3)-\frac{6}{\sqrt{2}}MC_2(x,x_1,x_2)C_1(x,x_3) \nonumber \\
&&-\frac{6}{\sqrt{2}}MC_1(x,x_1)C_2(x,x_2,x_3)-\frac{6}{\sqrt{2}}MC_2(x,x_1,x_3)C_1(x,x_2), \nonumber \\
&\vdots&.
\eea
For both the solutions, this is signaling the spontaneous breaking of conformal invariance. The Green function is then given by
\be
\label{eq:GomH}
C_1(p)=\frac{1}{p^2+2\mu_0^2},
\ee
and we have a single boson of mass $\sqrt{2}\mu_0$ with translation invariance never broken after quantization. We see that the classical theory displays a massive solution in any case, due to the nonlinear potential we obtained from the Starobinsky model.

\section{Quantum solution\label{sec4}}

From the analysis in the precedent section, the true minimum is located at
\be
\label{eq:G1_R2}
G_1=\phi_0=\frac{1}{6}(3M+\sqrt{3}\sqrt{3M^2+8R})\approx\sqrt{\frac{2}{3}R}=\mu_0,
\ee
in the considered approximation $3M^2<\mathcal{R}$, given the Dyson-Schwinger set of equations
\bea
&&\partial^2 G_1(x)-\mu_R^2G_1(x)+[G_1(x)]^3+G_3(x,x,x)=\Omega, \nonumber \\
&&\\ \nonumber
&&\partial^2G_2(x,y)-\mu_R^2G_2(x,y)+3[G_1(x)]^2G_2(x,y)=\delta^4(x-y), \nonumber \\
&&\\ \nonumber
&&\partial^2G_3(x,y,z)-\mu_R^2G_3(x,y,z)+3G_1^2(x)G_3(x,y,z)+6G_1(x)G_2(x,y)G_2(x,z) \\ \nonumber
&&+3G_2(x,z)G_3(x,x,y)+3G_2(x,y)G_3(x,x,z) \\ \nonumber
&&+3G_2(x,x)G_3(x,y,z)+3G_1(x)G_4(x,x,y,z)+G_5(x,x,x,y,z)=0, \\ \nonumber
&&\\ \nonumber
&&\partial^2G_4(x,y,z,w)-\mu_R^2G_4(x,y,z,w)+3G_1^2(x)G_4(x,y,z,w)+6G_2(x,y)G_2(x,z)G_2(x,w)\\ \nonumber
&&+6G_1(x)G_2(x,y)G_3(x,z,w)+6G_1(x)G_2(x,z)G_3(x,y,w)+6G_1(x)G_2(x,w)G_3(x,y,z)\\ \nonumber
&&+3G_2(x,y)G_4(x,x,z,w)+3G_2(x,z)G_4(x,x,y,w)  \\ \nonumber
&&+3G_2(x,w)G_4(x,x,y,z)+3G_2(x,x)G_4(x,y,z,w) \\ \nonumber
&&+3G_1(x)G_5(x,x,y,z,w)+G_6(x,x,x,y,z,w)=0, \\ \nonumber
&\vdots&,
\eea
where now we have set
\be
\label{eq:muR1}
\mu_R^2=\mu_0^2-3G_2(x,x),
\ee
with the correction due to quantum fluctuations evaluated in Appendix
B.
Thus, we have $\phi_0$ renormalized by quantum effects. 
This will yield for the 2P-correlation function
\be
\label{eq:G2_R2}
\partial^2G_2(x,y)+M_G^2G_2(x,y)=\delta^4(x-y),
\ee
where the final Mass Gap developed in the theory reads as (see Appendix B)
\begin{align}
M^2_G=2\mu_R^2 = \frac{8 \pi ^2 \left(\sqrt{3} \sqrt{3 M^2+8 R}+3 M\right)^2}{27 W\left(\frac{16}{27} \pi ^3 \left(\sqrt{3} \sqrt{3 M^2+8 R}+3 M\right)^2 e^{-1+\gamma +\frac{16 \pi ^2}{3}}\right)},
\end{align}
where $W(x)$ is the Lambert function that solves the equation $xe^x=w$.


\begin{figure}[H]\begin{center}
\includegraphics[scale=0.6]{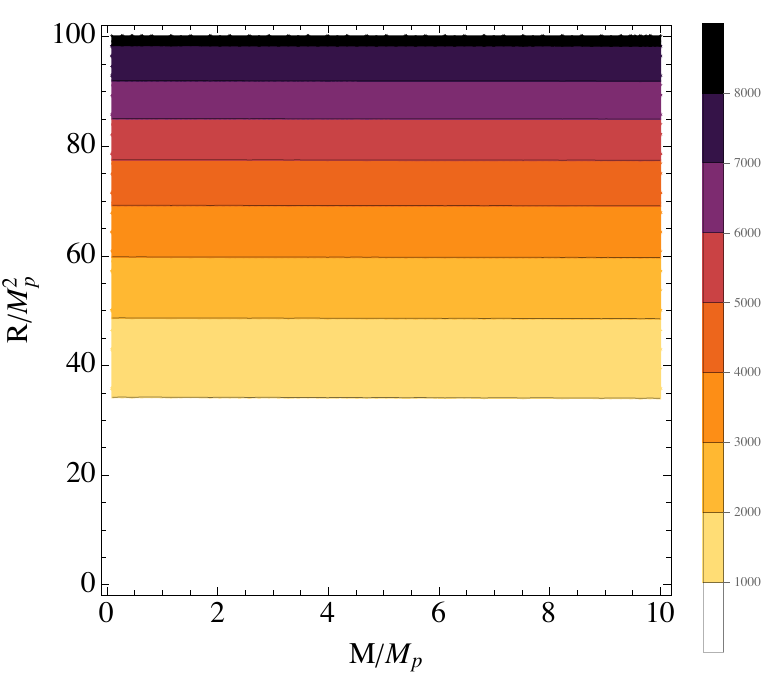}
\caption{ \it Plot of mass gap. \label{fig2}}
\end{center}
\end{figure}

We observe that the conformal factor that appears in passing from the Jordan to the Einstein frame, yields a propagating scalar field. The appearance of a mass gap in the quantum study this Starobinsky scalar field displays a breaking of the scale symmetry, reducing the geometry of the spacetime to that of the backbone geometry that has a null Ricci curvature. This shows a transition between a full blown quadratic gravity to a Einstein-Hilbert dominated gravity.

%
%
The mass gap is seen to be mostly independent on $M$. What really matters is the value of the Ricci scalar $\cal R$. Due to the condition ${\cal R}>3M^2$, there a large set of values (white region in Fig.\ref{fig2}) where the mass gap cannot exist. The meaning of the mass gap is that, due to the phase transition appearing in the UV-regime, the scalaron in the Starobinsky model becomes ineffective at lower energies,
confirming the analysis given above. 

\medskip

\section{Discussion and Conclusion\label{sec5}}



\begin{itemize}
\item We have proposed a simple model out of quadratic gravity simplified to the Einstein-Hilbert term plus a quadratic term of the Ricci scalar. This is the Starobinsky model (\ref{eq:star}). The model is derived assuming that the physics of interest, for strong fields, 
is described by a large and constant Ricci scalar granting the appearance of a mass gap in the Starobinsky scalaron field. This provides a transition scale from a quadratic dominated gravity to a Einstein-Hilbert dominated gravity.
\item The model is well represented through a single mode scalar field characterized by a potential (\ref{eq:Vfvd}) (see Fig.~\ref{fig1}) that can display the phenomenon of false vacuum decay, provided a condition is satisfied, already at classical level between 
the scalar Ricci curvature, that we assume constant, $\mathcal{R}$ satisfies the strong coupling condition $3M^2<\mathcal{R}$, being $M$ the mass term due to the quadratic term in the action (\ref{eq:star}).
\item  The scalar field can be studied and quantized in the limit of a strong field dynamics. For $\Lambda=0$, this yields a mass gap and a spectrum represented in eq.(\ref{eq:spe1}). This is a discrete spectrum with an infinite tower of harmonic-oscillator like excitations. The spectrum depends on $G_2(x,x)$, a constant to be renormalized. The mass gap scale is proportional to the Ricci scalar $\mathcal{R}$.
%
\item We quantize the model using the technique of Dyson-Schwinger equations described in Appendix B as it is amenable to exact analytical solutions for the 1P- and 2P-correlation functions providing a Gaussian solution for the quantum field theory.
\item The quantized theory provides as a solution for the metric, the 1P-correlation function given in eq.(\ref{eq:G1_R2}), 
%
%
and representing the vacuum of the theory. 
A mass gap is displayed
%
at the level of 2P-correlation function (\ref{eq:G2_R2}) that describes the propagating degrees of freedom.
\item The overall effect of a mass gap in the quadratic gravity makes it negligible at very low-energy where the EH term prevails.
\end{itemize}

Although we show the development of mass gap in the theory, if that were not the case, that is, we are not in the strong coupling regions and the conditions such as $\Lambda \gg M $ are not satisfied,
there would be no way to observe such unusual features: simply because the patches of the Universe where their effects are sizable are not compatible with observers. It is only in the nearly homogeneous and isotropic
Hubble patches that cosmic inflation may take place and experimental observations can eventually be done. This is one possible explicit quantum realization of the very popular Linde’s chaotic inflation \cite{Linde:1983gd} and can be understood at the full quantum level of the classical results of quadratic gravity (for details, see Ref.\cite{Salvio:2019ewf} ). This may naturally explain the near homogeneity and isotropy of our universe. We plan to study this using the mechanism presented in this paper and also extend this model adding matter content in future work. 

Similarly, an interesting area of application of our formalism is for Einstein-Gauss-Bonnet cosmological models \cite{Nojiri:2005jg,Cruz-DombrizS:2012,BenettiSCAL:2018,OdintsovOFF:2020,DeLaurentis:2015fea} that appear very promising in the understanding of the experimental data \cite{Odintsov:2025kyw}. We will address also this matter in future publications.

\medskip

\section*{Acknowledgement}

Authors thank Alexey Koshelev and Alberto Salvio for discussion.

\medskip

\section*{Appendix A: Derivation of Dyson-Schwinger equations for a quartic scalar field}\label{appB}

The technique we use to derive the set of Dyson-Schwinger equations in PDE shape has been devised by Bender, Milton and Savage \cite{Bender:1999ek} for a PT-invariant quantum mechanical model. We extended this approach to quantum field theory \cite{Frasca:2015yva}. For our convenience, we work with an Euclidean metric and write the partition function as
\be
Z[j]=\int[d\phi]e^{-\int d^4x[\frac{1}{2}(\partial\phi)^2-\frac{\lambda}{4}\phi^4+j\phi]}
\ee
where $j$ is arbitrary source and $\lambda$ the theory self-coupling. Our aim is to evaluate the full set of nP-correlation functions
\be
G_n(x_1,x_2,\ldots,x_n)=\left.\frac{\delta^n\ln Z}{\delta j(x_1)\delta j(x_2)\ldots\delta j(x_n)}\right|_{j=0},
\ee
so that we are able to get a complete solution for the quantum field theory and all the observables like decay rates and scattering cross sections will be computable through the LSZ theorem. It is straightforward to write down
\bea
G_1(x_1)&=&\left.\frac{1}{Z}\frac{\delta Z}{\delta j(x_1)}\right|_{j=0}, \nonumber \\
G_2(x_1,x_2)&=&\left.\frac{\delta G_1^{(j)}(x_1)}{\delta j(x_2)}\right|_{j=0}, \nonumber \\
G_3(x_1,x_2,x_3)&=&\left.\frac{\delta G_2^{(j)}(x_1,x_2)}{\delta j(x_3)}\right|_{j=0}, \nonumber \\
&\vdots&,
\eea
where
\be
G_n^{(j)}(x_1,x_2,\ldots,x_n)=\frac{\delta^n\ln Z}{\delta j(x_1)\delta j(x_2)\ldots\delta j(x_n)},
\ee
that are the correlation functions dependent on the source $j$. The equation of motion takes the form
\be
\partial^2\phi+\lambda\phi^3=j.
\ee
In order to get $G_1$, we average this equation with $Z$ obtaining
\be
\label{eq:G1j}
\partial^2G_1^{(j)}(x)+\lambda Z^{-1}\langle\phi^3\rangle=j,
\ee
but using
\be
Z G_1^{(j)}(x)=\langle\phi\rangle,
\ee
we can iterate the derivative with respect to $j(x)$ obtaining
\bea
&&Z G_2^{(j)}(x,x)+Z[G_1^{(j)}(x)]^2=\langle\phi^2\rangle, \nonumber \\
&&3Z G_1^{(j)}(x)G_2^{(j)}(x,x)+Z G_3^{(j)}(x,x,x)+Z[G_1^{(j)}(x)]^3=\langle\phi^3\rangle.
\eea
By direct substitution in eq.(\ref{eq:G1j}), we get the PDE for $G_1^{(j)}$ 
\be
\label{eq:G1j1}
\partial^2G_1^{(j)}(x)+\lambda \left\{3G_2^{(j)}(x,x)G_1^{(j)}(x)+G_3^{(j)}(x,x,x)+[G_1^{(j)}(x)]^3\right\}=j,
\ee
and, by setting $j=0$, one has
\be
\label{eq:G1j2}
\partial^2G_1(x)+\lambda [3G_1(x)G_2(x,x)+G_3(x,x,x)+G_1^3(x)]=0.
\ee
To get the equation for $G_2$. we need to derive eq.(\ref{eq:G1j1}) with respect to $j(x_2)$. This will yield
\be
\partial^2G_2^{(j)}(x,x_2)+\lambda\left\{3G_2^{(j)}(x,x)G_2^{(j)}(x,x_2)+3G_2^{(j)}(x,x)G_3^{(j)}(x,x,x_2)+G_4^{(j)}(x,x,x,x_2)+3[G_1^{(j)}(x)]^2G_2^{(j)}(x,x_2)\right\}.
\ee
By setting $j=0$, one gets the equation for $G_2$. The procedure can be iterated at any order to compute the higher-order correlation functions. This approach is successful when exact solution are known for $G_1$ and $G_2$. Only in this case, a Gaussian solution for the partition function can be obtained by selecting $G_n=0$ for $n>2$, provided at least two independent variables are identical.

\section*{Appendix B: Mass shift}\label{appC}

We want to evaluate the contribution of the mass shift in eq.(\ref{eq:muR1}) by evaluating the self-consistency equation
\be
\mu_R^2=\mu_0^2-3\int\frac{d^4p}{(2\pi)^4}\frac{1}{p^2+2\mu_R^2}.
\ee
Using dimensional regularization, we get
\be
\int\frac{d^Dp}{(2\pi)^4}\frac{1}{p^2+2\mu_R^2}=\frac{\Gamma(-1+\epsilon)}{(4\pi)^{2-\epsilon}}(\mu_R^2)^{1-\epsilon},
\ee
where we set $\epsilon=(4-D)/2$. Thus, doing an expansion for $\epsilon\rightarrow 0$, we get
\be
\mu_R^2=\mu_0^2-3\left[-\frac{\mu_R^2}{16\pi^2\epsilon}+\frac{\mu_R^2}{16\pi^2}\left(-1+\gamma+\log(\mu_R^2)+\log(4\pi)\right)\right]+O(\epsilon),
\ee
where $\gamma$ is the Euler-Mascheroni constant. We take care of the divergent part by considering it a correction to the coupling constant $M$. Thus, we can safely take the limit $\epsilon\rightarrow 0$ and we are left with
\be
\mu_R^2\left[1+\frac{3}{16\pi^2}\left(-1+\gamma+\log(\mu_R^2)+\log(4\pi)\right)\right]=\mu_0^2.
\ee
This equation can be solved using the Lambert function $W$, writing
\be
\mu_R^2=\frac{4 \pi ^2 \left(\sqrt{3} \sqrt{3 M^2+8 R}+3 M\right)^2}{27 W\left(\frac{16}{27} \pi ^3 \left(\sqrt{3} \sqrt{3 M^2+8 R}+3 M\right)^2 e^{-1+\gamma +\frac{16 \pi ^2}{3}}\right)}.
\ee


\medskip

\begin{thebibliography}{99}

\bibitem{LIGOScientific:2016aoc}
B.~P.~Abbott \textit{et al.} [LIGO Scientific and Virgo],
Phys. Rev. Lett. \textbf{116}, no.6, 061102 (2016)
doi:10.1103/PhysRevLett.116.061102
[arXiv:1602.03837 [gr-qc]].

\bibitem{EventHorizonTelescope:2019dse}
K.~Akiyama \textit{et al.} [Event Horizon Telescope],
Astrophys. J. Lett. \textbf{875}, L1 (2019)
doi:10.3847/2041-8213/ab0ec7
[arXiv:1906.11238 [astro-ph.GA]].

\bibitem{EventHorizonTelescope:2019ggy}
K.~Akiyama \textit{et al.} [Event Horizon Telescope],
Astrophys. J. Lett. \textbf{875}, no.1, L6 (2019)
doi:10.3847/2041-8213/ab1141
[arXiv:1906.11243 [astro-ph.GA]].

\bibitem{EventHorizonTelescope:2022xnr}
K.~Akiyama \textit{et al.} [Event Horizon Telescope],
Astrophys. J. Lett. \textbf{930}, no.2, L12 (2022)
doi:10.3847/2041-8213/ac6674
[arXiv:2311.08680 [astro-ph.HE]].

\bibitem{EventHorizonTelescope:2022wok}
K.~Akiyama \textit{et al.} [Event Horizon Telescope],
Astrophys. J. Lett. \textbf{930}, no.2, L14 (2022)
doi:10.3847/2041-8213/ac6429
[arXiv:2311.09479 [astro-ph.HE]].

\bibitem{EventHorizonTelescope:2022xqj}
K.~Akiyama \textit{et al.} [Event Horizon Telescope],
Astrophys. J. Lett. \textbf{930}, no.2, L17 (2022)
doi:10.3847/2041-8213/ac6756
[arXiv:2311.09484 [astro-ph.HE]].

\bibitem{Goroff:1985sz}
M.~H.~Goroff and A.~Sagnotti,
Phys. Lett. B \textbf{160}, 81-86 (1985)
doi:10.1016/0370-2693(85)91470-4

\bibitem{Goroff:1985th}
M.~H.~Goroff and A.~Sagnotti,
Nucl. Phys. B \textbf{266}, 709-736 (1986)
doi:10.1016/0550-3213(86)90193-8

\bibitem{Salvio:2018crh}
A.~Salvio,
Front. in Phys. \textbf{6}, 77 (2018)
doi:10.3389/fphy.2018.00077
[arXiv:1804.09944 [hep-th]].

\bibitem{Salvio:2024joi}
A.~Salvio,
JCAP \textbf{07}, 092 (2024)
doi:10.1088/1475-7516/2024/07/092
[arXiv:2404.08034 [hep-th]].

\bibitem{Donoghue:2021cza}
J.~F.~Donoghue and G.~Menezes,
Nuovo Cim. C \textbf{45}, no.2, 26 (2022)
doi:10.1393/ncc/i2022-22026-7
[arXiv:2112.01974 [hep-th]].

\bibitem{Weinberg:1974tw}
S.~Weinberg,
PRINT-74-1313 (HARVARD).

\bibitem{Deser:1975nv}
S.~Deser,
Conf. Proc. C \textbf{750926}, 229-254 (1975)

\bibitem{Stelle:1976gc}
K.~S.~Stelle,
Phys. Rev. D \textbf{16}, 953-969 (1977)
doi:10.1103/PhysRevD.16.953

\bibitem{Barvinsky:2017zlx}
A.~O.~Barvinsky, D.~Blas, M.~Herrero-Valea, S.~M.~Sibiryakov and C.~F.~Steinwachs,
JHEP \textbf{07}, 035 (2018)
doi:10.1007/JHEP07(2018)035
[arXiv:1705.03480 [hep-th]].

\bibitem{Gibbons:1994cg}
G.~W.~Gibbons and S.~W.~Hawking,

\bibitem{Menotti:1989ms}
P.~Menotti,
Nucl. Phys. B Proc. Suppl. \textbf{17}, 29-38 (1990)
doi:10.1016/0920-5632(90)90218-J

\bibitem{Buchbinder-Lyahovich:1987}
I.~L.~Buchbinder and S.~L.~Lyahovich,
``Canonical quantisation and local measure of $R^2$ gravity"
 Class. Quantum Grav. \textbf{4} (1987)  1487
doi:10.1088/0264-9381/4/6/008

\bibitem{Buchbinder:1991ne}
I.~L.~Buchbinder, I.~Y.~Karataeva and S.~L.~Lyakhovich,
Class. Quant. Grav. \textbf{8}, 1113-1125 (1991)
doi:10.1088/0264-9381/8/6/010

\bibitem{Buchbinder:1992rb}
I.~L.~Buchbinder, S.~D.~Odintsov and I.~L.~Shapiro,
``Effective Action in Quantum Gravity'',
(Routledge, New York, 1992). 
doi:10.1201/9780203758922

\bibitem{Salvio:2015gsi}
A.~Salvio and A.~Strumia,
Eur. Phys. J. C \textbf{76}, no.4, 227 (2016)
doi:10.1140/epjc/s10052-016-4079-8
[arXiv:1512.01237 [hep-th]].


 \bibitem{Pais}
A.~Pais and G.~E.~Uhlenbeck,  ``On Field theories with nonlocalized action'',  Phys.\ Rev.\  {79} (1950) 145.

\bibitem{Pauli}  W.~Pauli, ``On Dirac's New Method of Field Quantization", Rev.\ Mod.\ Phys.\ 15 (1943) 175
doi:10.1103/RevModPhys.15.175 
      
\bibitem{Dirac}    P.~A.~M.~Dirac, ``The physical interpretation of quantum mechanics," 
	Proc.\ R.\ Soc.\ Lond.\ A 180, 1 (1942)
	doi:10.1098/rspa.1942.0023
	


 \bibitem{ostro}
M. Ostrogradsky, ``Memoires sur les \'equations diff\'erentielles relatives au probl\`eme des isop\'erim\`etres," Mem. Ac. St. Petersbourg VI (1850) 385. \href{https://babel.hathitrust.org/cgi/pt?id=mdp.39015038710128;view=1up;seq=405}{Pdf available online}.


\bibitem{Wheeler}
 J. A. Wheeler, in Battelle Rencontres, edited by C. DeWitt and J. A. Wheeler (Benjamin, New York, 1968).


\bibitem{Strumia:2017dvt}
A.~Strumia,
MDPI Physics \textbf{1}, no.1, 17-32 (2019)
doi:10.3390/physics1010003
[arXiv:1709.04925 [quant-ph]].

\bibitem{Salvio:2019wcp}
A.~Salvio,
Eur. Phys. J. C \textbf{79}, no.9, 750 (2019)
doi:10.1140/epjc/s10052-019-7267-5
[arXiv:1907.00983 [hep-ph]].

\bibitem{Salvio:2020axm}
A.~Salvio,
Int. J. Mod. Phys. A \textbf{36}, no.08n09, 2130006 (2021)
doi:10.1142/S0217751X21300064
[arXiv:2012.11608 [hep-th]].

\bibitem{Bender:2008gh}
C.~M.~Bender and P.~D.~Mannheim,
Phys. Rev. D \textbf{78}, 025022 (2008)
doi:10.1103/PhysRevD.78.025022
[arXiv:0804.4190 [hep-th]].

\bibitem{Bender:2007wu}
C.~M.~Bender and P.~D.~Mannheim,
Phys. Rev. Lett. \textbf{100}, 110402 (2008)
doi:10.1103/PhysRevLett.100.110402
[arXiv:0706.0207 [hep-th]].

\bibitem{Bender:2023cem}
C.~M.~Bender and D.~W.~Hook,
[arXiv:2312.17386 [quant-ph]].

\bibitem{Salvio:2017xul}
A.~Salvio,
Eur. Phys. J. C \textbf{77}, no.4, 267 (2017)
doi:10.1140/epjc/s10052-017-4825-6
[arXiv:1703.08012 [astro-ph.CO]].

\bibitem{Gross:2020tph}
C.~Gross, A.~Strumia, D.~Teresi and M.~Zirilli,
Phys. Rev. D \textbf{103}, no.11, 115025 (2021)
doi:10.1103/PhysRevD.103.115025
[arXiv:2007.05541 [hep-th]].

\bibitem{Held:2021pht}
A.~Held and H.~Lim,
Phys. Rev. D \textbf{104}, no.8, 084075 (2021)
doi:10.1103/PhysRevD.104.084075
[arXiv:2104.04010 [gr-qc]].

\bibitem{Held:2023aap}
A.~Held and H.~Lim,
Phys. Rev. D \textbf{108}, no.10, 104025 (2023)
doi:10.1103/PhysRevD.108.104025
[arXiv:2306.04725 [gr-qc]].

\bibitem{Deffayet:2021nnt}
C.~Deffayet, S.~Mukohyama and A.~Vikman,
Phys. Rev. Lett. \textbf{128}, no.4, 041301 (2022)
doi:10.1103/PhysRevLett.128.041301
[arXiv:2108.06294 [gr-qc]].

\bibitem{Deffayet:2023wdg}
C.~Deffayet, A.~Held, S.~Mukohyama and A.~Vikman,
JCAP \textbf{11}, 031 (2023)
doi:10.1088/1475-7516/2023/11/031
[arXiv:2305.09631 [gr-qc]].

\bibitem{Lee:1969fy}
T.~D.~Lee and G.~C.~Wick,
Nucl. Phys. B \textbf{9}, 209-243 (1969)
doi:10.1016/0550-3213(69)90098-4

\bibitem{Salvio:2018kwh}
A.~Salvio, A.~Strumia and H.~Veerm\"ae,
Eur. Phys. J. C \textbf{78}, no.10, 842 (2018)
doi:10.1140/epjc/s10052-018-6311-1
[arXiv:1808.07883 [hep-th]].

\bibitem{Donoghue:2019fcb}
J.~F.~Donoghue and G.~Menezes,
Phys. Rev. D \textbf{100}, no.10, 105006 (2019)
doi:10.1103/PhysRevD.100.105006
[arXiv:1908.02416 [hep-th]].

\bibitem{Planck2018:inflation}
Y.~Akrami, et~al., {Planck 2018 results. X. Constraints on inflation}, Astron.
  Astrophys. 641 (2020) A10
  doi:10.1051/0004-6361/201833887
 [\hhref{arXiv:1807.06211}].


 \bibitem{WeinbergAS} S. Weinberg, in Understanding the Fundamental Constituents of Matter, ed. A. Zichichi (Plenum 	Press, New York, 1977). S. Weinberg,  in General  Relativity: An  Einstein Centenary Survey, 	edited by S. W. Hawking and W. Israel (Cambridge University Press, 1980) pp. 790-831.


\bibitem{Holdom:2021hlo}
B.~Holdom,
Phys. Rev. D \textbf{105}, no.4, 046008 (2022)
doi:10.1103/PhysRevD.105.046008
[arXiv:2107.01727 [hep-th]].

\bibitem{Anselmi:2017ygm}
D.~Anselmi,
JHEP \textbf{06}, 086 (2017)
doi:10.1007/JHEP06(2017)086
[arXiv:1704.07728 [hep-th]].

\bibitem{Frasca:2020jbe}
M.~Frasca and A.~Ghoshal,
Class. Quant. Grav. \textbf{38}, no.17, 17 (2021)
doi:10.1088/1361-6382/ac161b
[arXiv:2011.10586 [hep-th]].

\bibitem{Frasca:2020ojd}
M.~Frasca and A.~Ghoshal,
JHEP \textbf{21}, 226 (2020)
doi:10.1007/JHEP07(2021)226
[arXiv:2102.10665 [hep-th]].

\bibitem{Frasca:2021iip}
M.~Frasca, A.~Ghoshal and N.~Okada,
Phys. Rev. D \textbf{104}, no.9, 096010 (2021)
doi:10.1103/PhysRevD.104.096010
[arXiv:2106.07629 [hep-th]].

\bibitem{Frasca:2022duz}
M.~Frasca, A.~Ghoshal and A.~S.~Koshelev,
Class. Quant. Grav. \textbf{41}, no.1, 015014 (2024)
doi:10.1088/1361-6382/ad0a51
[arXiv:2202.09578 [hep-ph]].

\bibitem{Frasca:2022gdz}
M.~Frasca, A.~Ghoshal and A.~S.~Koshelev,
Phys. Lett. B \textbf{841}, 137924 (2023)
doi:10.1016/j.physletb.2023.137924
[arXiv:2207.06394 [hep-th]].

\bibitem{Bender:1999ek}
C.~M.~Bender, K.~A.~Milton and V.~M.~Savage,
Phys. Rev. D \textbf{62}, 085001 (2000)
doi:10.1103/PhysRevD.62.085001
[arXiv:hep-th/9907045 [hep-th]].

\bibitem{Frasca:2015yva}
M.~Frasca,
Eur. Phys. J. Plus \textbf{132}, no.1, 38 (2017)
[erratum: Eur. Phys. J. Plus \textbf{132}, no.5, 242 (2017)]
doi:10.1140/epjp/i2017-11321-4
[arXiv:1509.05292 [math-ph]].

\bibitem{Chaichian:2018cyv}
M.~Chaichian and M.~Frasca,
Phys. Lett. B \textbf{781}, 33-39 (2018)
[arXiv:1801.09873 [hep-th]].

\bibitem{Frasca:2022pjf}
M.~Frasca, A.~Ghoshal and S.~Groote,
Nucl. Part. Phys. Proc. \textbf{324-329}, 85-89 (2023)
doi:10.1016/j.nuclphysbps.2023.01.019
[arXiv:2210.02701 [hep-ph]].

\bibitem{Frasca:2021yuu}
M.~Frasca, A.~Ghoshal and S.~Groote,
Phys. Rev. D \textbf{104}, no.11, 114036 (2021)
doi:10.1103/PhysRevD.104.114036
[arXiv:2109.05041 [hep-ph]].

\bibitem{Frasca:2021mhi}
M.~Frasca, A.~Ghoshal and S.~Groote,
Nucl. Part. Phys. Proc. \textbf{318-323}, 138-141 (2022)
doi:10.1016/j.nuclphysbps.2022.09.029
[arXiv:2109.06465 [hep-ph]].

\bibitem{Frasca:2022lwp}
M.~Frasca, A.~Ghoshal and S.~Groote,
Phys. Lett. B \textbf{846}, 138209 (2023)
doi:10.1016/j.physletb.2023.138209
[arXiv:2202.14023 [hep-ph]].


\bibitem{Frasca:2015wva}
M.~Frasca,
Eur. Phys. J. Plus \textbf{131}, no.6, 199 (2016)
doi:10.1140/epjp/i2016-16199-x
[arXiv:1504.02299 [hep-ph]].

\bibitem{Frasca:2019ysi}
M.~Frasca,
Eur. Phys. J. C \textbf{80}, no.8, 707 (2020)
doi:10.1140/epjc/s10052-020-8261-7
[arXiv:1901.08124 [hep-ph]].

\bibitem{Frasca:2016sky}
M.~Frasca,
Eur. Phys. J. C \textbf{77}, no.4, 255 (2017)
doi:10.1140/epjc/s10052-017-4824-7
[arXiv:1611.08182 [hep-th]].



\bibitem{Frasca:2012ne}
M.~Frasca,
J. Nonlin. Math. Phys. \textbf{20}, no.4, 464-468 (2013)
doi:10.1080/14029251.2013.868256
[arXiv:1212.1822 [hep-th]].



\bibitem{Frasca:2010ce}
M.~Frasca,
PoS \textbf{FACESQCD}, 039 (2010)
doi:10.22323/1.117.0039
[arXiv:1011.3643 [hep-th]].

\bibitem{Frasca:2008tg}
M.~Frasca,
Nucl. Phys. B Proc. Suppl. \textbf{186}, 260-263 (2009)
doi:10.1016/j.nuclphysbps.2008.12.058
[arXiv:0807.4299 [hep-ph]].

\bibitem{Frasca:2009yp}
M.~Frasca,
Mod. Phys. Lett. A \textbf{24}, 2425-2432 (2009)
doi:10.1142/S021773230903165X
[arXiv:0903.2357 [math-ph]].

\bibitem{Frasca:2008zp}
M.~Frasca,
Int. J. Mod. Phys. E \textbf{18}, 693-703 (2009)
doi:10.1142/S0218301309012781
[arXiv:0803.0319 [hep-th]].

\bibitem{Frasca:2007uz}
M.~Frasca,
Phys. Lett. B \textbf{670}, 73-77 (2008)
doi:10.1016/j.physletb.2008.10.022
[arXiv:0709.2042 [hep-th]].

\bibitem{Frasca:2006yx}
M.~Frasca,
Int. J. Mod. Phys. A \textbf{22}, 2433-2439 (2007)
doi:10.1142/S0217751X07036427
[arXiv:hep-th/0611276 [hep-th]].

\bibitem{Frasca:2005sx}
M.~Frasca,
Phys. Rev. D \textbf{73}, 027701 (2006)
[erratum: Phys. Rev. D \textbf{73}, 049902 (2006)]
doi:10.1103/PhysRevD.73.049902
[arXiv:hep-th/0511068 [hep-th]].



\bibitem{Frasca:2017slg}
M.~Frasca,
Nucl. Part. Phys. Proc. \textbf{294-296}, 124-128 (2018)
doi:10.1016/j.nuclphysbps.2018.02.005
[arXiv:1708.06184 [hep-ph]].

\bibitem{Frasca:2005mv}
M.~Frasca,
Int. J. Mod. Phys. A \textbf{22}, 1441-1450 (2007)
doi:10.1142/S0217751X07036282
[arXiv:hep-th/0509125 [hep-th]].

\bibitem{Frasca:2005fs}
M.~Frasca,
Int. J. Mod. Phys. D \textbf{15}, 1373-1386 (2006)
doi:10.1142/S0218271806009091
[arXiv:hep-th/0508246 [hep-th]].

\bibitem{Frasca:2009bc}
M.~Frasca,
J. Nonlin. Math. Phys. \textbf{18}, no.2, 291-297 (2011)
doi:10.1142/S1402925111001441
[arXiv:0907.4053 [math-ph]].

\bibitem{Frasca:2013tma}
M.~Frasca,
Eur. Phys. J. C \textbf{74}, 2929 (2014)
doi:10.1140/epjc/s10052-014-2929-9
[arXiv:1306.6530 [hep-ph]].

\bibitem{Frasca:2023qii}
M.~Frasca and A.~Ghoshal,
Eur. Phys. J. C \textbf{84}, no.10, 1101 (2024)
doi:10.1140/epjc/s10052-024-13458-2
[arXiv:2306.17818 [hep-th]].

\bibitem{Frasca:2023eoj}
M.~Frasca and A.~Ghoshal,
[arXiv:2311.15258 [hep-ph]].

\bibitem{Frasca:2022kfy}
M.~Frasca, A.~Ghoshal and N.~Okada,
J. Phys. G \textbf{51}, no.3, 035001 (2024)
doi:10.1088/1361-6471/ad170e
[arXiv:2201.12267 [hep-th]].

\bibitem{Calcagni:2022tls}
G.~Calcagni, M.~Frasca and A.~Ghoshal,
[arXiv:2206.09965 [hep-th]].

\bibitem{Calcagni:2022gac}
G.~Calcagni, M.~Frasca and A.~Ghoshal,
Int. J. Mod. Phys. D \textbf{33}, no.01, 2350111 (2024)
doi:10.1142/S0218271823501110
[arXiv:2211.06957 [hep-th]].

\bibitem{Frasca:2022vvp}
M.~Frasca, A.~Ghoshal and A.~S.~Koshelev,
Eur. Phys. J. C \textbf{82}, no.12, 1108 (2022)
doi:10.1140/epjc/s10052-022-11057-7
[arXiv:2203.15020 [hep-th]].



\bibitem{Frasca:2024fuq}
M.~Frasca, A.~Ghoshal and N.~Okada,
[arXiv:2408.00093 [hep-ph]].

\bibitem{Chatterjee:2024dgw}
A.~Chatterjee, M.~Frasca, A.~Ghoshal and S.~Groote,
[arXiv:2407.21179 [hep-ph]].

\bibitem{Frasca:2024pmv}
M.~Frasca, A.~Ghoshal and N.~Okada,
[arXiv:2402.12462 [hep-ph]].

\bibitem{Frasca:2024ame}
M.~Frasca, A.~Ghoshal and N.~Okada,
[arXiv:2410.00135 [hep-ph]].


\bibitem{Hartle:1983ai}
J.~B.~Hartle and S.~W.~Hawking,
Phys. Rev. D \textbf{28}, 2960-2975 (1983)
doi:10.1103/PhysRevD.28.2960

\bibitem{DeWitt:1967yk}
B.~S.~DeWitt,
Phys. Rev. \textbf{160}, 1113-1148 (1967)
doi:10.1103/PhysRev.160.1113

\bibitem{Starobinsky:1980te}
A.~A.~Starobinsky,
Phys. Lett. B \textbf{91}, 99-102 (1980)
doi:10.1016/0370-2693(80)90670-X

\bibitem{Ade:2015lrj}
P.~A.~R.~Ade \textit{et al.} [Planck],
Astron. Astrophys. \textbf{594}, A20 (2016)
doi:10.1051/0004-6361/201525898
[arXiv:1502.02114 [astro-ph.CO]].

\bibitem{BICEP:2021xfz}
P.~A.~R.~Ade \textit{et al.} [BICEP and Keck],
Phys. Rev. Lett. \textbf{127}, no.15, 151301 (2021)
doi:10.1103/PhysRevLett.127.151301
[arXiv:2110.00483 [astro-ph.CO]].

\bibitem{Benedetti:2009rx}
D.~Benedetti, P.~F.~Machado and F.~Saueressig,
Mod. Phys. Lett. A \textbf{24}, 2233-2241 (2009)
doi:10.1142/S0217732309031521
[arXiv:0901.2984 [hep-th]].

\bibitem{Falls:2020qhj}
K.~Falls, N.~Ohta and R.~Percacci,
Phys. Lett. B \textbf{810}, 135773 (2020)
doi:10.1016/j.physletb.2020.135773
[arXiv:2004.04126 [hep-th]].

\bibitem{Klevansky:1992qe}
S.~P.~Klevansky,
Rev. Mod. Phys. \textbf{64}, 649-708 (1992)
doi:10.1103/RevModPhys.64.649



\bibitem{Mukhanov:1981xt}
V.~F.~Mukhanov and G.~V.~Chibisov,
JETP Lett. \textbf{33}, 532-535 (1981)

\bibitem{Kaneda:2010ut}
S.~Kaneda, S.~V.~Ketov and N.~Watanabe,
Mod. Phys. Lett. A \textbf{25}, 2753-2762 (2010)
doi:10.1142/S0217732310033918
[arXiv:1001.5118 [hep-th]].




\bibitem{Whitt:1984pd}
B.~Whitt,
Phys. Lett. B \textbf{145}, 176-178 (1984)
doi:10.1016/0370-2693(84)90332-0

\bibitem{Kehagias:2013mya}
A.~Kehagias, A.~Moradinezhad Dizgah and A.~Riotto,
Phys. Rev. D \textbf{89}, no.4, 043527 (2014)
doi:10.1103/PhysRevD.89.043527
[arXiv:1312.1155 [hep-th]].





















































































































\bibitem{Linde:1983gd}
A.~D.~Linde,
Phys. Lett. B \textbf{129}, 177-181 (1983)
doi:10.1016/0370-2693(83)90837-7



\bibitem{Salvio:2019ewf}
A.~Salvio,
Phys. Rev. D \textbf{99}, no.10, 103507 (2019)
doi:10.1103/PhysRevD.99.103507
[arXiv:1902.09557 [gr-qc]].




\bibitem{Nojiri:2005jg}
  S.~Nojiri and S.~D.~Odintsov,
  Phys.\ Lett.\ B {\bf 631} (2005) 1, 
  [hep-th/0508049].

\bibitem{Cruz-DombrizS:2012}
 A. de la Cruz-Dombriz and D. Saez-Gomez,
 Class. Quant. Grav. 29 (2012) 245014, [arXiv:1112.4481].

\bibitem{BenettiSCAL:2018}
 M. Benetti, S. Santos da Costa, S. Capozziello, J.S. Alcaniz and M. De Laurentis,
 Int. J. Mod. Phys. D 27 (2018) 1850084, arXiv:1803.00895

  \bibitem{OdintsovOFF:2020}
 S.D. Odintsov, V.K. Oikonomou, F.P. Fronimos and K.V. Fasoulakos,
 Phys. Rev. D 102 (2020) 104042, [arXiv:2010.13580]

\bibitem{DeLaurentis:2015fea}
  M.~De Laurentis, M.~Paolella and S.~Capozziello,
  Phys.\ Rev.\ D {\bf 91} (2015) no.8,  083531
  doi:10.1103/PhysRevD.91.083531
  [arXiv:1503.04659 [gr-qc]].

\bibitem{Odintsov:2025kyw}
S.~D.~Odintsov, V.~K.~Oikonomou and G.~S.~Sharov,
[arXiv:2503.17946 [gr-qc]].

\end{thebibliography}
\end{document}